# A Spiking Neural Network based on Neural Manifold for Augmenting Intracortical Brain-Computer Interface Data

Shengjie Zheng[1,2][0000-0002-9370-8141], Wenyi Li[1,2], Lang Qian[3], Chenggang He[2] and Xiaojian Li[2(✉)]

[1] University of Chinese Academy of Sciences, Beijing, China
`zhengshengjie20@mails.ucas.edu.cn`
[2] Brain Cognition and Brain Disease Institute (BCBDI), Shenzhen-Hong Kong Institute of Brain Science, Shenzhen Institute of Advanced Technology, Chinese Academy of Sciences, Shenzhen, China
`{ sj.zheng, wy.li, xj.li } @siat.ac.cn`
[3] Tsinghua Shenzhen International Graduate School, Tsinghua University, Shenzhen, China
`ql20@mails.tsinghua.edu.cn`

**Abstract.** Brain-computer interfaces (BCIs), transform neural signals in the brain into instructions to control external devices. However, obtaining sufficient training data is difficult as well as limited. With the advent of advanced machine learning methods, the capability of brain-computer interfaces has been enhanced like never before, however, these methods require a large amount of data for training and thus require data augmentation of the limited data available. Here, we use spiking neural networks (SNN) as data generators. It is touted as the next-generation neural network and is considered as one of the algorithms oriented to general artificial intelligence because it borrows the neural information processing from biological neurons. We use the SNN to generate neural spike information that is bio-interpretable and conforms to the intrinsic patterns in the original neural data. Experiments show that the model can directly synthesize new spike trains, which in turn improves the generalization ability of the BCI decoder. Both the input and output of the spiking neural model are spike information, which is a brain-inspired intelligence approach that can be better integrated with BCI in the future.

## 1  Introduction

Motor control is a very important aspect of human life, and humans interact through a variety of behaviors. However, motor abilities often receive limitations such as stroke, amyotrophic lateral sclerosis (ALS), or other injuries or neurological disorders that disrupt the neural pathways connecting the brain to the rest of the body, leading to paralysis[1-3]. A motor-brain-computer interface is a system that helps patients by recording neural signals from internal areas of the brain and decoding them into control instructions, as shown in Fig. 1. Intracortical brain-computer interface systems require microelectrode arrays that can be implanted in the cerebral cortex for long periods of time to record the activity of dozens to hundreds of neurons.

In clinical trials, implantable brain-computer interface systems have not yet reached a level of performance that can widely assist patients with severe paralysis. In addition, it is unclear whether current brain-computer interface methods can be effectively used continuously for long periods of time. For example, after the brain is implanted with microelectrodes, the microelectrodes' ability to capture neural information decreases over time. This is because the electrodes cause partial damage to the brain, triggering the growth of glial scar tissue around the electrodes, blocking the electrode contacts from the neurons. At the same time, electrode contacts gradually break down in the electrolyte solution environment in the brain, reducing the electrode's electrical sensing performance. For decoding algorithms, the performance of the decoder also decreases over time. Neuroelectrodes will shift due to brain shaking, resulting in changes in the recorded information, and most decoding algorithms cannot adapt to such changes in neural information. Neuroplasticity (the learning process of neurons) leads to the reconnection of neurons as well as changes in the strength of the connections. Therefore, the neural information of the brain changes in response to the learning process. Therefore, the current BCI decoders need to be recalibrated periodically and require a large amount of neural information to obtain good performance. Also, the decoder trained based on neural information is not universal, and different subjects need to collect different neural information to train the decoder. Therefore, the training data obtained in clinical trials are very limited.

Here we will explore the data deficiency from the perspective of cortical motor neuron populations. Recent theoretical and experimental work hypothesizes that neural function is based on specific patterns of neural population activity rather than on individual neurons[4]. The neural population dynamics exist in low-dimensional neural manifolds in a high-dimensional neural space[5]. Here, we employ a bio-interpretive SNN that mimics the neural information generation as well as the communication of biological neural populations. We analyze motor cortical neural population data recorded from monkeys to derive motor-related neural population dynamics. The neural spike properties of the SNN itself allow the direct generation of biologically meaningful spike trains that match the activity of real biological neural populations. We explored the interaction between the spike train synthesizer and the BCI decoder. Our results show that based on a small amount of training data as a template, data conforming to the dynamics of neural populations are generated, thus enhancing the decoding ability of the BCI decoder.

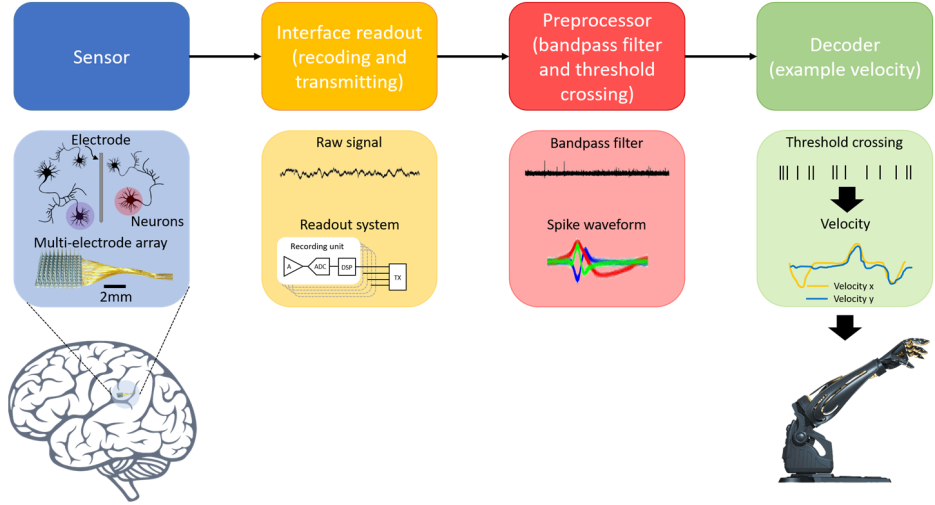

**Fig. 1.** The neural information is recorded using an array of electrodes, each measuring nearby neural activity. The raw neural signal is transformed into a digital signal by the signal acquisition system. The signals can be processed by band-pass filters and threshold crossing for spike detection. Finally, a decoder is used to model the neural spike signal with motor control and the control signal is sent to the prosthesis for control. Image modified from [1].

## 2    Background and Related Work

### 2.1    Spiking Neuron

The Spiking Neuron Model simulates the membrane potential changes in biological neurons that result in action potentials (spikes). These spikes are generated by the neuronal soma and transmitted to other neurons along axons and synapses. Spiking neuronal models can be divided into various categories, the most detailed being the biophysical model based on ion channel simulation (Hodgkin-Huxley Model), which describes the change in membrane potential as a function of input current and the activation of an ion channel. In terms of mathematical calculations, a simpler one is the Leaky Integrate and fire model (LIF model), which describes the membrane potential transformation as a function of the input current and the LIF model as the basic unit of the SNN, which is expressed as follows (Equation 1) and (Equation 2).

$$\tau_m \frac{du_i}{dt} = -(u_i - u_{rest}) + RI_i \quad (1)$$

$$\frac{dI_i}{dt} = -\frac{I_i(t)}{\tau_s} + \sum_j W_{ij} S_j(t) \quad (2)$$

where $u_i$ is the membrane potential of the neuron, $\tau_m$ is the membrane time constant, $u_{rest}$ is the resting potential of the membrane, $R$ is the membrane resistance, $I_i$ is the input current, $\tau_s$ is the synaptic time constant, and $W_{ij}$ is the synaptic connection strength.

When the neuronal membrane potential $u_i$ exceeds the threshold $\vartheta$, the neuron firing as well as resets the membrane potential to $u_{\text{rest}}$ ($u_{\text{rest}} < \vartheta$). Incorporating the reset property, it follows that

$$\frac{du_i}{dt} == -\frac{1}{\tau_m}(u_i - u_{\text{rest}}) + RI_i + S_i(t)(u_{\text{rest}} - \vartheta) \tag{3}$$

$$S_i(t) = \sum_k \delta(t - t_i^k) \tag{4}$$

$S_i(t)$ is the spike train emitted by neuron $i$ (represented by the sum of the Dirac delta equation, $\delta$), and $t_i^k$ is the spike firing time of the neuron at time $k^{th}$.

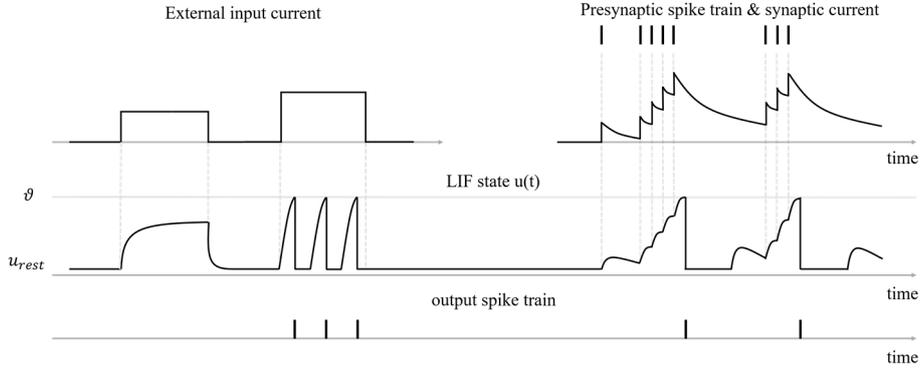

**Fig. 2.** The membrane potential change caused by the input current to a single LIF spiking neuron. When the membrane potential reaches a threshold, a spike is released and the membrane potential will be reset. Image modified from [6].

## 2.2    Neural Manifold

Recent technological developments have provided tools to detect the activity of large numbers of neurons, as well as computational statistics and modeling tools to analyze the activity of neuronal populations. The number of neurons that can be recorded simultaneously is currently about a few hundred, which is far less than the number of neurons contained in the corresponding cortex of the brain. Nevertheless, the current motor brain-computer interfaces are achieving good results. This is because in recent theories and studies it is believed that neural function is built on the activation of specific population-wide activity patterns-neural mode-rather than on the independent modulation of individual neurons[4, 7]. These neural modes define the neural manifold, which is a plane that captures most of the variance in recorded neural activity, as shown in Fig. 3.

Assuming that network connectivity constrains the possible patterns of population activity, the population dynamics will not explore the full high-dimensional neural space but will instead remain confined to a low-dimensional surface within the full space[8]. In the simplest case, the neural manifold is flat, as in the hyperplane in Fig.3 and the plane is spanned by two neural modes u1 and u2.

Meanwhile, recent research points out that this time-dependent activation of the neural modes as their latent dynamics. In this theory, the activity of each recorded neuron expresses a weighted combination of the latent dynamics from all the neural modes. Neural modes can be estimated empirically from the recorded activity by applying dimensionality reduction techniques such as principal component analysis (PCA) to construct a low-dimensional manifold embedded in the empirical neural space spanned by the recorded neurons. Thus, this intrinsic neural manifold can exhibit to some extent the neuronal population dynamics pattern.

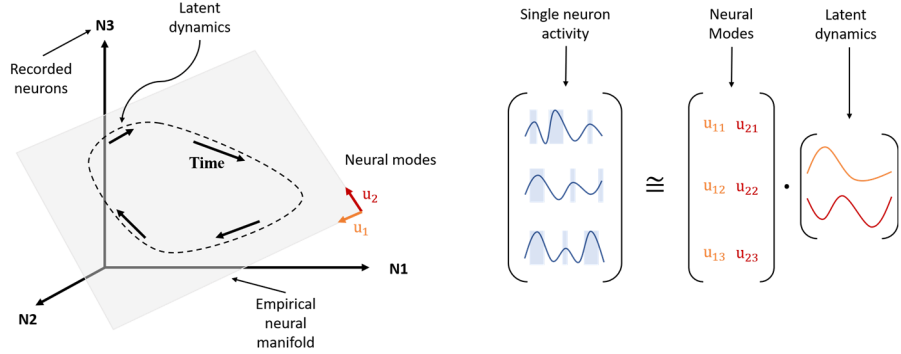

**Fig. 3.** The activity of recorded neurons (N1, N2 and N3) is represented within the empirical neural space, with each axis representing a recorded neuron. Recent studies suggest that neuronal activity may be a weighted combination of the latent dynamics of neural modes. Image modified from [8].

### 2.3    Neural Decoding

Neural decoding uses the activity recorded by the brain to predict variables in the external world. These predictions derived from decoding can be used to control devices (e.g., robotic arms) or to better understand how brain regions relate to the external world. Thus, neural decoding is a core tool for neural engineering and neural data analysis. In essence, neural decoding is a regression (or classification) problem that relates neural signals to specific variables. Decoding of neural signals has been better achieved using modern machine learning techniques.

For decoders, the aim is to understand the information contained in the neural activity or to understand how the information in the neural activity is related to external variables. Recurrent neural networks based on Simple RNN, Gated Recurrent Unit (GRU), Long Short-Term Memory Network (LSTM), and Non-recurrent Decoders such as Wiener Filter, Kalman Filter, XGBoost, Feedforward Neural Network (FNN), Naïve Bayes, Support Vector Regression (SVR) are often used[9]. For any decoder, the purpose is to predict the recorded output value based on the neural signal, as shown in Fig. 4.

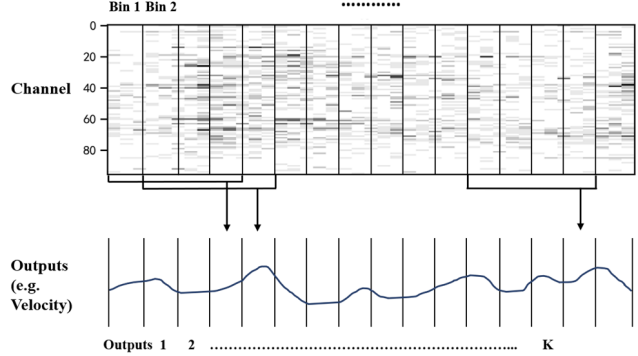

**Fig. 4.** Use the decoder to predict the output corresponding to the channel binned data for a specific time bin.

### 2.4    Related work

In brain-computer interfaces systems, it has not been possible to explore universally applicable hidden dynamics from the neural information due to the complexity within the neural system. In recent years, there has been the use of advanced machine learning techniques to explore the interaction between spike generators and BCI decoders. For example, a data-driven generative adversarial network (GAN) based on data from multiple tasks trains the spike train generator, and the model can be used to synthesize new data quickly[10].

A recent approach, SNN_EEG[11], can generate EEG data by SNN, and the generated spike trains are converted into EEG data as templates by filters. Our approach complements this work in the following aspects. First, the SNN_EEG aims to generate EEG data, and the SNN was fitted as a regression to the EEG template. In contrast, we generate spike data, constrained in a neural population dynamics manner, to generate biologically consistent data.

## 3    Method

### 3.1    Overview Architecture

The whole data generation process is divided into four steps, as shown in **Fig. 5**. The first step is to process and dimensionality reduction of the neural spike data collected from the brain-computer interface of the same task trial to derive the neural population dynamics in low-dimensional neural manifold space, and the second step is to build an SNN model with the real neuron population dynamics manifold as the target function for supervised learning. In the third step, during the learning process, the spike trains generated by the SNN are subjected to principal component analysis (PCA) for dimensionality reduction, and the resulting activity laws need to be approximated by the real neuron population dynamics. In the fourth step, after the network iteration, a perturbation neuron is set in the SNN, and the perturbation neuron is responsible for generating noisy pulses, thus generating noisy and rule-based neural spike data.

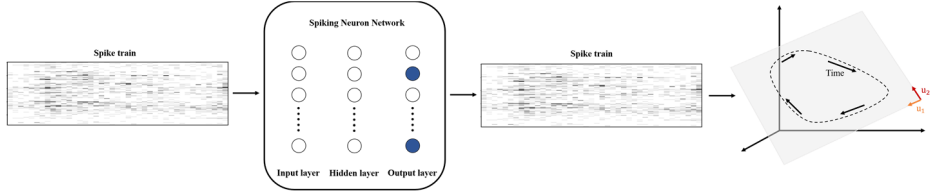

**Fig. 5.** The SNN generates spike trains data and the dynamics of the spiking neurons in the output layer are in the neural manifold space.

A forward SNN is used to generate the spike trains, each node is a LIF neuron model, and the network architecture consists of three layers, the input layer consists of a fixed sequence of Poisson generated spike trains with a frequency of 10 Hz as input, followed by the hidden layer and the output layer. During the training process, the spiking neurons in the output layer generate the corresponding spike trains according to the input, and each neuron in the output layer is processed with a sliding window to transform it into firing rate, followed by dimensionality reduction, so as to map the spiking neurons in the output layer into a low-dimensional neural manifold. The neural manifold activity pattern of the real neuron population in the same dimension is used as the objective function, and the mean squared difference (MSE) is used as the loss function. The training results need to satisfy that the activity pattern of the spiking neuron population in the output layer needs to approximate the activity pattern of the real neuron population.

After the iteration of the network, some neurons in the output layer of the SNN are activated, and Poisson neurons are used as the basis and set as perturbation neurons. A random spike signal based on Poisson distribution is input to the perturbation neuron, which causes the perturbation neuron to generate a noisy signal. The output layer can thus generate neural spike information data with noise and in accordance with the real biological neural population dynamics.

### 3.2    Surrogate Gradient Descent

The learning algorithm is based on a supervised learning algorithm that uses an algorithm called surrogate gradient descent to make the weights of the SNN to be updated[12].

$$S_i[n] \propto \Theta(u_i[n] - u_{\text{threshold}}) \quad (5)$$

$S_i[n]$ is denoted as spike trains, where $\Theta(x)$ is denoted as a Heaviside step function, due to the non-differentiable nature of the spike signal. In back propagation, the $\Theta(x)$ function is replaced by $\sigma(x)$. Then we have $S_i[n] \propto \sigma(u_i[n] - u_{\text{threshold}})$, where $\sigma(x) = 1/1 + \exp(-x)$.

### 3.3	Generating spike trains dataset

After the network iteration is completed, some neurons in the output layer are set as perturbation neurons. A Poisson process-based spike train is input into the perturbation neuron, which causes the neuron to generate noisy spike trains.

Implemented within the program, the calculation can be simplified, and the probability that the spike is issued within the time step $\Delta t$ can be considered as $r\Delta t$, where $r$ is the firing rate and $x_{\text{rand}}$ is a random variable with values within 0 to 1.

$$r_i \Delta t \begin{cases} > x_{\text{rand}} & \text{firing} \\ \leq x_{\text{rand}} & \text{none} \end{cases} \tag{6}$$

## 4	Experiments

### 4.1	Data Processing and Analysis

We will use a dataset from Shenoy's lab, Stanford University[1], which includes a 96-channel Utah array recording the motor cortex of a non-human primate performing a center-out reaching task[13], as shown in Fig.6. The data contain spike trains recorded in each channel, as well as the cursor positions displayed on the computer screen by the monkey control, and the cursor velocities while recording 8 different trial types.

The spikes trains in the channel, as well as the position, are based on a 1ms window, by downsampling to a 25ms window. At the same time, the spike trains are converted to firing rate to improve the signal-to-noise ratio, and Gaussian smoothing is performed on the cursor position information to remove part of the noisy data.

The channel-based spike trains in the data are obtained by threshold crossing and not by spike sorting, which refers to the conversion of the acquired raw neural signal into a spike train for each neuron, based on the similarity of an action potential. In a recent study, it was shown that the neural manifold built based on the spike information of threshold crossing is similar to that of spike sorting[14]. Therefore, we choose the neural manifold built based on threshold crossing.

In the data analysis, the neural information of multiple same trials is averaged to obtain the average neural information of the same trial. Subsequently, the average neural information is dimensionalized to obtain the neural manifold corresponding to the low-dimensional neural space, as shown in Fig.7.

---

[1]	https://github.com/slinderman/stats320

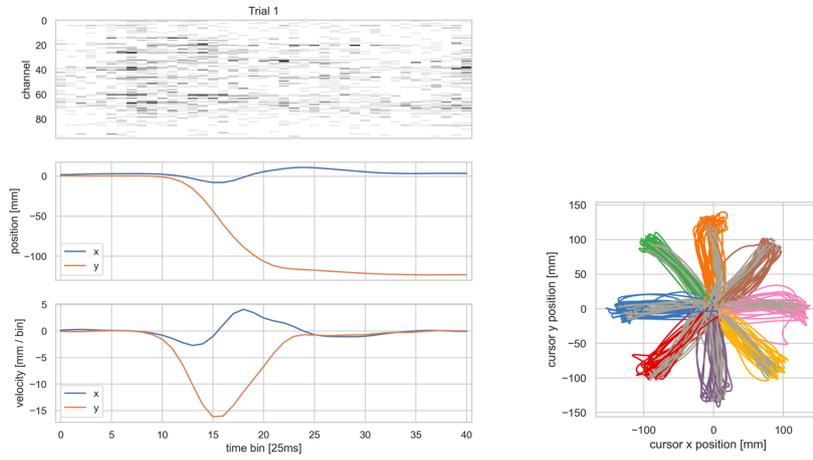

**Fig. 6.** BCI Data that has been processed

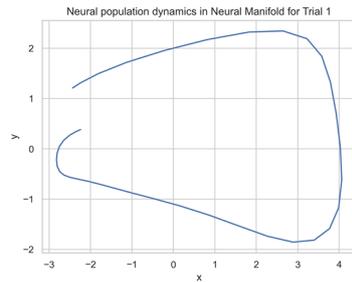

**Fig. 7.** The neuronal population dynamics in neural manifold for Trial 1 (Gaussian smoothing is done)

### 4.2   Experimental Setting

The SNN consists of three layers, each with 100, 400, and 100 LIF neurons, respectively, with fixed Poisson spike trains at 10 Hz as the input, followed by a hidden layer and an output layer. During the training process, the spiking neurons in the output layer generate the corresponding spike trains according to the input, and the sliding window processing is performed for each neuron of the output layer to transform it into firing rate data, followed by dimensionality reduction, thus mapping the spiking neurons in the output layer into the low-dimensional neural manifold. The neural manifold activity pattern of the real neuron population in the same dimension is used as the objective function, and the MSE (mean squared error) is used as the loss function. The training results need to satisfy that the activity pattern of the population of spiking neurons in the output layer needs to approximate the activity pattern of the population of real neurons.

After the network is trained, the model parameter that minimizes the loss value is selected, and for this experiment epoch 2130 is selected, and the loss value of this epoch is minimized, as shown in Fig.8. After that, the generation of data is performed, and 1~3% of neurons in the output layer are started as perturbation neurons to generate noise information. After that, the trained SNN converts the fixed input spike trains into output spike trains. This is repeated several times to generate the spike train dataset corresponding to the brain-computer interface experiment Trial 1. In this experiment, we generated data with similar neural dynamics as the original data Trial 1, as shown in Fig.8.

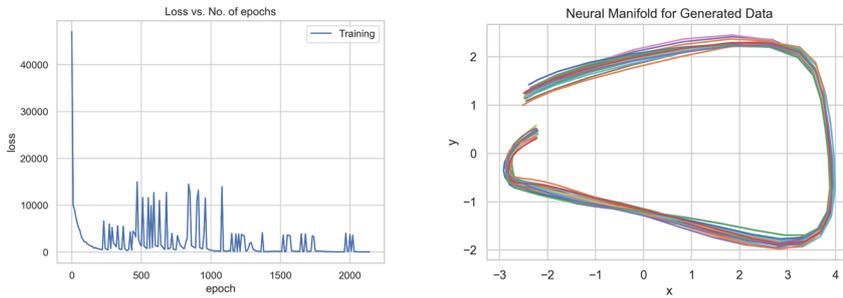

**Fig. 8** The training process (Left). The neuronal population dynamics in neural manifold for generated data (Right).

After the iteration, the inputs and outputs of the model are shown in **Fig.9**. The inputs to the model are always fixed and the network model training process transforms the fixed inputs into biologically meaningful spike trains. These output spike trains are sparser compared to the Poisson input.

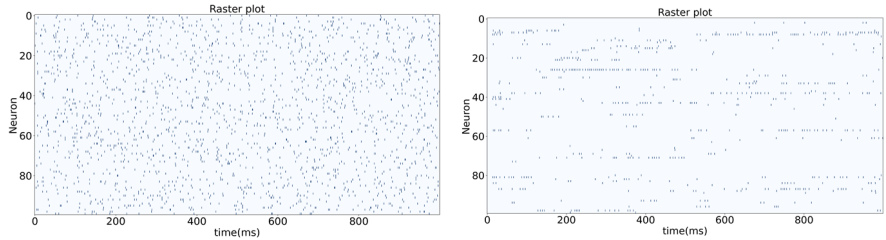

**Fig. 9.** Raster plot comparison, input spike trains and output spike trains for 100 neurons.

### 4.3   Results

We decode the generated data in combination with the original Trial1 data. Performance evaluation is performed using decoders LSTM, RNN, and FNN. Based on the original brain-computer interface experimental data combined with the generated neural spike signals, the recorded cursor movement speed values were predicted and the performance evaluation metrics were used $R^2$.

1. Expanded dataset based on the original dataset (96 channels) to train the decoder to derive the cursor movement x and y.
2. Expanded dataset based on part of the original dataset (10 channels) to train the decoder to derive the cursor movement x and y.
3. Meanwhile, the decoding effect of the expanding data is compared with the decoding effect of the original data set.

Table 1. Decoding results on the different datasets

| Method | LSTM | | RNN | | FNN | |
|---|---|---|---|---|---|---|
| | Validate | Test | Validate | Test | Validate | Test |
| | x | x | x | x | x | x |
| | y | y | y | y | y | y |
| 96 channels | 0.790 | 0.900 | 0.651 | 0.666 | 0.735 | 0.874 |
| | 0.600 | 0.868 | 0.740 | 0.767 | 0.793 | 0.889 |
| 10 channels | 0.695 | 0.752 | 0.536 | 0.593 | 0.404 | 0.667 |
| | 0.487 | 0.610 | 0.375 | 0.406 | 0.365 | 0.530 |
| 96 channels and 100 neurons | **0.887** | **0.915** | **0.749** | **0.799** | **0.789** | **0.912** |
| | **0.870** | **0.916** | **0.819** | **0.813** | **0.883** | **0.917** |
| 10 channels and 100 neurons | **0.752** | **0.856** | **0.735** | **0.847** | **0.659** | **0.921** |
| | **0.814** | **0.830** | **0.828** | **0.818** | **0.845** | **0.849** |

Comparing the results using $R^2$ as an evaluation metric, as shown in Table 1. In our experiments, we generated spike trains corresponding to 100 spiking neurons, and we can observe that the best results can be obtained in the complete original data combined with the generated data. In addition, even only a small amount of original data combined with the generated dataset can be obtained with good results. The results show that the network is able to generate spike trains with similar neural dynamics as the original data based on the neural manifold space as a constraint and is able to enhance the decoding ability in the corresponding decoding experiments while extending the original dataset. In our model can be suitable to be applied in clinical trials, the network model can reduce the reliance on the original data due to the limited daily data collection.

## 5    Conclusion

Intracortical brain-machine interfaces take neural information from the motor cortex in the cerebral cortex, and the neural information available in clinical experiments is often limited, and the decoder needs data to be retrained for fine-tuning over time. The goal of this work is to provide a method to combine an intracortical brain-computer interface with brain-inspired intelligence as a spike generator capable of generating neural population dynamics corresponding to the Trial. In the future, with the development of neuromorphic chips, SNN can be more effective with lower energy consumption and can eventually help BCI effectively.